\documentstyle[11pt, epsf]{article}

\catcode`@=11  
\@addtoreset{equation}{section}
\def\@eqnnum{{\rm \theequation}}
\renewcommand{\theequation}{(\arabic{section}.\arabic{equation})}               

\renewcommand{\Re}{I\!\!R}                                                                              

\begin{document}         


\title{\bf Einstein's equation and geometric asymptotics.\footnote
{To appear in the proceedings of the 15th International
Conference on General Relativity and Gravitation (GR 15), Pune,
India, December 16 - 21, 1997.}}      
\author {Helmut Friedrich\\
Albert-Einstein-Institut\\ Max-Planck-Institut f\"ur Gravitationsphysik\\
Schlaatzweg 1, 14473 Potsdam, Germany }
\date{}
\maketitle

\section{Introduction}

The intimate relations between Einstein's equation, conformal
geometry, geometric asymptotics, and the idea of an isolated system
in general relativity have been pointed out by Penrose
\cite{penrose:let}, \cite{penrose:scri} many years ago. A detailed
analysis of the interplay of conformal geometry with Einstein's
equation (\cite{friedrich:1981a}, \cite{friedrich:1981b}) allowed us to
deduce from the conformal properties of the field equations a method to
derive under various assumptions definite statements about the
feasibility of the idea of geometric asymptotics. 

More recent investigations have demonstrated the possibility to analyse
the most delicate problem of the subject -- the behaviour of
asymptotically flat solutions to  Einstein's equation in the region
where ``null infinity meets space-like infinity'' -- to an arbitrary
precision. Moreover, we see now
that the, initially quite abstract, analysis yields methods for dealing
with practical issues. Numerical calculations of complete space-times in
finite grids without cut-offs become feasible now. Finally, already at
this stage it is seen that the completion of these investigations will
lead to a clarification and deeper understanding of the idea of an
isolated system in Einstein's theory of gravitation. In the following I
wish to give a survey of the circle of ideas outlined above, 
emphasizing the interdependence of the structures and the naturalness of
the concepts involved. 

\section{Geometric asymptotics and conformal field equations}

To illustrate the notion of geometric asymptotics I shall discuss the  
content of the following theorem (\cite{friedrich:first},
\cite{friedrich:global}).

\vspace{.3cm}

{\em The set of smooth asymptotically simple solutions is open in the 
set of all smooth, maximal, globally hyperbolic solutions to Einstein's
equation
\begin{equation}
\label{einstlneg}
Ric[\tilde{g}] = \lambda\,\tilde{g},\,\,\,\,\,\,\lambda < 0,\,\,\,\,\,
\,\,\,\,\,\,\,sign(\tilde{g}) = (1, -1, -1, -1)
\end{equation}
with compact space sections (``Nonlinear stability of asymptotic
simplicity, case  $\lambda < 0$'').}

\vspace{.3cm}

To see what is entailed by the asserted asymptotic simplicity of the
solutions, we recall its technical definition \cite{penrose:let},
\cite{penrose:scri} in a form convenient for our purpose (cf. Figure
$1$). A space-time $(\tilde{M}, \tilde{g})$ is called asymptotically
simple iff:

\begin{enumerate}

\item
All null geodesics $\gamma(t)$ in $(\tilde{M}, \tilde{g})$ are complete.

\item
There exists a smooth function $\Omega > 0$ on $\tilde{M}$ with
$\Omega(\gamma(t)) \rightarrow 0$ as $|t| \rightarrow \infty$,

\item
there exists an extension of $\tilde{M}$ to a smooth manifold 
$M = \tilde{M} \cup {\cal J}$ with boundary 
${\cal J} = {\cal J}^{-} \cup \,{\cal J}^{+}$, where
${\cal J}^{-} \cap \,{\cal J}^{+} = \emptyset$, such that:

\item
$\Omega$ extends smoothly to $M$ with: $\Omega = 0$, $d\,\Omega \neq 0$
on ${\cal J}$,

\item
$g = \Omega^2\,\tilde{g}$ extends to a smooth Lorentz metric on $M$,

\item
Each null geodesic acquires an endpoint in the past on ${\cal J}^{-}$
and an endpoint in the future on ${\cal J}^{+}$. 

\end{enumerate}

This definition implies in particular that the space-times considered in
the theorem above are complete with ${\cal J}^{-}$ (${\cal J}^{+}$)
being a space-like hypersurface which represents the infinite null and
time-like past (future). The fact that we can associate with the
``physical'' space-time $(\tilde{M}, \tilde{g})$ the smooth compact
conformal extension $(M, g)$ (which is unique apart from conformal
diffeomorphisms) implies among other things that the conformal Weyl
tensor $C^{\mu}\,_{\nu \lambda \rho}$ of $g$, which agrees on
$\tilde{M}$ with that of $\tilde{g}$, vanishes on ${\cal J}$. Thus
asymptotic simplicity characterizes the asymptotic behaviour of the
physical space-time. This characterization is optimal; any attempt to
strengthen it further would remove the generality of the admissible
solutions. Being determined solely in terms of the conformal structure
-- certainly the most important substructure of the metric in general
relativity -- this characterization is certainly as geometrical as we
could wish.

When we shall consider later on solutions for different signs of the
cosmological constant $\lambda$ we shall assume the definition
above in approprately modified form. In particular, it may be necessary
to drop the first condition and the decomposition of the hypersurface
${\cal J}$ into two components. Since all this has been discussed at
length in the literature I shall not dwell on it any further.         

\begin{figure}
\centerline{\epsffile{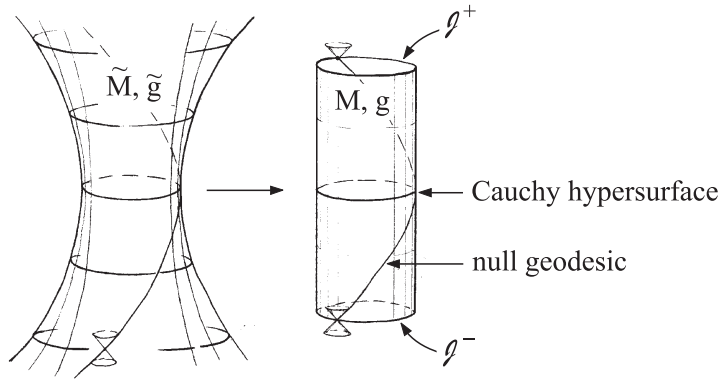}} 
\centering
\begin{minipage}[b]{11cm}
\let \normalsize = \footnotesize
\caption{\sl Embedding the physical space-time into its
conformal extension.}
\end{minipage}
\end{figure}

Pick now any solution of the type considered in the theorem -- the
standard example is given by De Sitter space -- and consider the
Cauchy data induced on an arbitrary smooth Cauchy hypersurface of it. 
If we change these data by a finite but sufficiently small amount, the
theorem tells us that these data develop again into a solution
satisfying the definition above. Thus the completeness as well as the
specific fall-off behaviour of the gravitational field will be
preserved. There exist further results which show that this stability
property is retained when Einstein's equations are coupled to
conformally well behaved fields like Yang-Mills fields. A reliable
estimation of the size of the changes in the data which are admissible
here is at present not available. Explicit examples make it clear that we
need to impose restrictions.

We emphasize that the asymptotic behaviour of the perturbed solutions
asserted in the theorem is implied solely by the choice of data and by
the specific properties of Einstein's equation exhibited below, it is
not imposed artificially from outside.

As has been stressed already in \cite{penrose:let}, \cite{penrose:scri},
the property of asymptotic simplicity is of interest for:

{\em Practical reasons: In many considerations complicated
limits in the physical space-time $(\tilde{M}, \tilde{g})$ can be
replaced by simple differential geometric calculations in the conformal
space-time $(M, g)$.}

{\em Conceptual reasons: Physical notions based on
approximations in the physical space-time can be replaced by notions
which are defined in a precise way in terms of fields induced on the
conformal boundary ${\cal J}$.} 

After the idea had been introduced, the further study of asymptotic
simplicity concentrated for a long time mainly on these two aspects (cf.
the surveys \cite{ashtekar:1984}, \cite{geroch}). In view of the theorem
above we can now add:

{\em Asymptotic simplicity is a natural concept for solutions to}
\[
Ric[\tilde{g}] = \lambda\,\tilde{g},\,\,\,\,\,\,\lambda < 0,\,\,\,\,\,
\,\,\,\,\,\,\,sign(\tilde{g}) = (1, -1, -1, -1).
\]

We have seen that the asymptotically simple solutions form islands, not
singular peaks in the sea of solutions. What is lying in between?
Largely uncharted waters. To survey the islands or to explore in any
generality what can be found on and beyond their coast lines, we would
need to develop new tools. Of course, asymptotically simple space-times
have also been studied by the techniques of ``exact solutions''. Explicit
solutions can locate the islands or allow us to learn about new features
of gravitational fields by analysing solutions lying in between. There
exist e.g. explicit solutions to \ref{einstlneg} which develop pieces
of a smooth conformal boundary, horizons, and singularities. However, we
cannot directly obtain statements about general classes of solutions in
this way. 

Since ``existence'' appears to have different meanings for different
members of the relativity community, I wish to point out here that in the
theorem above I do not simply talk about presumptive solutions of which
the first few coefficients in a formal expansion have been determined.
Results of the generality considered above are necessarily based on
abstract arguments to establish the existence of solutions.
Nevertheless, the solutions referred to in this context have the
same mathematical ``reality'' and precision as explicit solutions. I
shall have no further reason to consider the field of exact
solutions and refer the interested reader to \cite{bicak} for a survey.

\vspace{.3cm} 

We do not only know by abstract arguments that the space-times
considered in the theorem have the very specific asymptotic
behaviour asserted there but we can also pinpoint the structural
origin of it. Using the fields 
$g = \Omega^2 \tilde{g}$ and $\Omega$, and a frame 
$e_k = e^{\mu}\,_k\,\partial_{x^{\mu}}$ satisfying 
$g(e_i, e_k) \equiv g_{ik} = diag(1, -1, -1, -1)$, we can derive from
Einstein's equation the ``conformal field equations''
\cite{friedrich:1981a}, \cite{friedrich:1981b} 
for the unknowns  
\[
e^{\mu}\,_k,\,\,\,\,\,
\Gamma_i\,^j\,_k,\,\,\,\,\,
d^{i}\,_{jkl} = \Omega^{-1}\,C^{i}\,_{jkl},\,\,\,\,\,
L_{ij} = \frac{1}{2}\,R_{ij} - \frac{1}{12}\,R \,g_{ij},
\]
\[
\Omega,\,\,\,\,\,
\bar{s} = \frac{1}{4}\nabla_{i}\nabla^{i}\Omega
+ \frac{1}{24}\,R\,\Omega,
\]
where $\Gamma_i\,^j\,_k$ denotes the connection coefficients in the frame
$e_k$ and $C^{i}\,_{jkl}$, $R_{ij}$, $R$ the conformal Weyl tensor, the
Ricci tensor, and the Ricci scalar of $g$ respectively. The conformal
field equations are given by
\[
[e_{p},e_{q}] =
(\Gamma_{p}\,^{l}\,_{q} - \Gamma_{q}\,^{l}\,_{p})\,e_{l},
\]
\[
e_{p}(\Gamma_{q}\,^{i}\,_{j}) - e_{q}(\Gamma_{p}\,^{i}\,_{j}) 
- 2\,\Gamma_{k}\,^{i}\,_{j}\,\Gamma_{[p}\,^{k}\,_{q]} 
+ 2\,\Gamma_{[p}\,^{i}\,_{|k|} \Gamma_{q]}\,^{k}\,_{j} 
\]
\[
= 2\,g^{i}\,_{[p}\, L_{q]j} - 2\,g^{ik}\,g_{j[p}\,L_{q]k}
+ \Omega \, d^{i}\,_{jpq}, 
\]
\[
\nabla_{i} d^{i}\,_{jkl} = 0,
\]
\[
\nabla_{i} L_{jk} - \nabla_{j} L_{ik}
= \nabla_{l} \Omega \, d^{l}\,_{kij},
\]
\[
\nabla_{i}\, \nabla_{j} \Omega = - \Omega \, L_{ij} 
+ \bar{s} g_{ij}, 
\]
\[
\nabla_{i} \bar{s} = - L_{ij} \nabla^{j} \Omega,
\]
\[
6 \Omega\,\bar{s} - 3\,\nabla_{j}\Omega\,\nabla^{j}\Omega = \lambda. 
\]
The derivation of these equations can be found in the literature and I
just point out a few of their properties. In regions where $\Omega > 0$
the system is equivalent to Einstein's equation. For the special
choice $\Omega \equiv 1$ the last four equations become trivial
identities and the first three equations reduce to the system which
forms the basis of the Newman-Penrose spin frame formalism 
\cite{newman:penrose:I}. The most important feature of the above system,
and in fact the structural background of our results, is the observation
that:

{\em The conformal field equations are, in a suitable sense,
hyperbolic. This is true irrespective of the sign of $\Omega$.}

Two important things are coming together here. The conformal field
equations do not contain terms of the form $\Omega^{-1}$, which occur
when the Ricci tensor of $g$ is expressed in terms of $\Omega$ and $g$.
Secondly, after a suitable choice of gauge conditions the largely
overdetermined system above implies symmetric hyperbolic systems of
propagation equations which are such that they preserve under their
evolution law the constraints which are implied by the system as well. 
It should be noted that the gauge conditions need to include a condition
which determines the conformal factor. A way to impose such a condition
is to prescribe the Ricci scalar $R$ of $g$ as a given function on the
manifold $M$. The existence results which led to our theorem are based
on the hyperbolicity of the conformal field equations.

\section{Time-like conformal boundaries}

If an asymptotically simple space-time solves Einstein's equation
\begin{equation}
\label{einstl} 
\tilde{R}_{\mu \nu} = \lambda\,\tilde{g}_{\mu \nu} 
\end{equation}
near ${\cal J}$, the function $\Omega$ satisfies  
$g^{\mu \nu}\,\nabla_{\mu}\,\Omega\,\nabla_{\nu}\,\Omega
= - \frac{1}{3}\,\lambda$ on ${\cal J}$, i.e. the sign of the
cosmological constant determines the causal properties of the conformal
boundary \cite{penrose:scri}. Therefore, if we want to assess the
richness of the class of asymptotically simple solutions to Einstein's
equation with $\lambda > 0$, for which  ${\cal J}$ is time-like, we need
to analyse initial boundary value problems where initial data are given
on a space-like slice and boundary data are prescribed on the (time-like)
conformal boundary ${\cal J}$ at space-like and null infinity
(cf. Figure $2$). In terms of the physical space-time this is an unsual
initial boundary value problem. The boundary data are to be given on a
boundary at space-like and null infinity to be associated with a
space-time which is not available yet and which itself is to be
determined partly by the boundary data.

\begin{figure}
\centerline{\epsffile{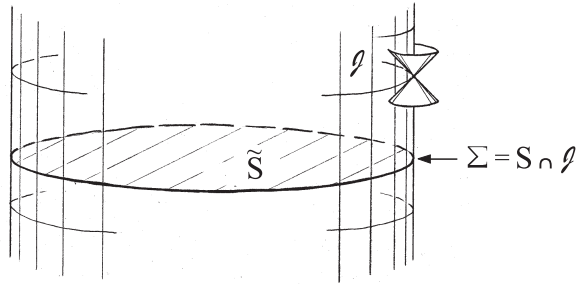}} 
\centering
\begin{minipage}[b]{11cm}
\let \normalsize = \footnotesize
\caption{\sl The conformal extension in the case $\lambda > 0$.} 
\end{minipage}
\end{figure}

Nevertheless, this initial boundary value problem can be analysed  
in detail. Let $\lambda > 0$ be given. Data are prescribed on smooth
3-manifolds $\tilde{S} = S \setminus \Sigma$ and ${\cal J}$, where $S$
is an orientable compact manifold with boundary $\Sigma$, 
${\cal J} = \Re \times \Sigma$, and we assume $\Sigma$ to be identified
with the subset $\{0\} \times \Sigma$ of ${\cal J}$. The data are given
as follows:

{\em On $\tilde{S}$ we prescribe smooth, asymptotically simple standard
Cauchy data $\tilde{h}_{\mu \nu}$, $\tilde{\chi}_{\mu \nu}$, which
satisfy on $\tilde{S}$ the $\lambda$-constraints for space-like
hypersurfaces.}

Though it should be obvious, we shall explain below the
notion of ``asymptotic simplicity'' for initial data sets more
carefully. It restricts the fall-off behaviour of the data near infinity.
In the case of the anti-De Sitter covering space, which is the standard
example of the situation considered in this chapter, an initial data
set of this type is given by $\tilde{S} = \Re^3$, $\tilde{\chi}_{\mu
\nu} = 0$, and a metric $\tilde{h}_{\mu \nu}$ such that 
$( \tilde{S}, - \tilde{h}_{\mu \nu})$ is a simply connected, complete
Riemannian space of negative constant curvature.   

{\em On ${\cal J}$ we prescribe as boundary data a smooth, 3-dimensional
Lorentzian conformal structure for which the slice $\Sigma$ is
space-like.}

If an initial boundary value problem has a smooth solution,
the initial and the boundary data are not quite independent. Thus,
without going into details, we require:

{\em The data on $S$ and ${\cal J}$ satisfy the
``compatibility conditions'' at $\Sigma$ which are implied by the
conformal field equations.}  

Given smooth data on $S$, there is no problem to construct data on
${\cal J}$ such that the compatibility conditions are satisfied.

We set $M = \Re \times S$, perform the obvious identifications
${\cal J} \simeq \partial M$, 
$\tilde{S} \hookrightarrow S \simeq \{0\} \times S \subset M$, and 
write  $\tilde{M} = M \setminus {\cal J}$. By $t$ we denote the
function on $M$, resp. on $\tilde{M}$, which is induced by the
projection $M = \Re \times S \rightarrow \Re$. Finally, let $\Omega$ be a
smooth function on $M$ with $\Omega > 0$ on $\tilde{M}$, $d\,\Omega \neq
0$ on ${\cal J}$. The function it induces on $\tilde{M}$ will also be
denoted by $\Omega$. We can state now the following theorem
\cite{friedrich:I}.  

{\em For given data on $\tilde{S}$ and ${\cal J}$ as described above
there is a $T > 0$ such that on  
$\tilde{M}' = \{p \in \tilde{M}|\,|t(p)| < T\} $ there exists a unique
(up to diffeomorphisms) smooth solution $\tilde{g}$ to  
$\tilde{R}_{\mu \nu} = \lambda\,\tilde{g}_{\mu\nu}$ which induces the
given initial data on $\tilde{S}$ and which is such that  
$g = \Omega^2\,\tilde{g}$ extends to a smooth Lorentz metric on   
$M' = \{p \in M|\,|t(p)| < T\}$ which induces the conformal structure
(up to diffeomorphisms) prescribed on $M' \cap {\cal J}$.} 

We have given here a characterization of all smooth asymptotically simple
solutions with $\lambda > 0$. Our result is global in space in the strong
sense that not only the space-like but also the null geodesics which
run out to $M' \cap {\cal J}$ are complete in that direction. It is
only local in time and in fact I did not try to show more; the cases 
$\lambda > 0$ and $\lambda \le 0$ are distinguished by a fundamental
difference in their global causal structures. This statement can be made
precise \cite{friedrich:I} in terms of invariants of the conformal
structure.

It is a remarkable feature of the initial boundary value problem above
that the data on the boundary can be prescribed in a simple
covariant way. This is not the case in the ``finite'' initial
boundary value problem for Einstein's vacuum field equations where the
boundary is thought of as a hypersurface in a smooth vacuum space-time 
\cite{friedrich:nagy}. Nevertheless, apart from the additional freedom
which allows us to characterize in the finite problem the location of
the boundary, the freedom to prescribe data is the same in both
problems. This indicates again the naturalness of the geometric
characterization of ``infinity'' which is inherent in the definition of
asymptotic simplicity.

There arise subtleties from the non-compactness of the initial
hypersurface $\tilde{S}$. To explain this we need to be
more specific about the nature of our data. In the context of initial
data ``asymptotic simplicity'' means the following: (i) The Riemannian
space $(\tilde{S}, - \tilde{h}_{\mu \nu})$ is conformally compactifiable
to a smooth Riemannian space $(S, - h_{\mu \nu})$ for which $\Sigma$
represents the conformal boundary at infinity. (ii) The data 
$\tilde{h}_{\mu \nu}$, $\tilde{\chi}_{\mu \nu}$ are transformed by the
appropriate conformal transformation laws into conformal initial
data $h_{\mu \nu}$, $\chi_{\mu \nu}$ which extend smoothly to $\Sigma$. 
(iii) The conformal initial data set $(S, h_{\mu\nu}, \chi_{\mu\nu})$
satisfies those fall-off conditions at $\Sigma$ which are implied by the
requirement that it can be embedded smoothly into an asymptotically
simple solution to \ref{einstl} as a hypersurface which has intersection
$\Sigma = S \cap {\cal J}$ with the conformal boundary and which is
space-like everywhere.

The existence of large classes of ``free data'' which can be used to
construct data  $h_{\mu \nu}$, $\chi_{\mu \nu}$ satisfying (i) to (iii)
follows from \cite{acf}, \cite{ac}, \cite{kannar}. We notice that
condition (iii) implies among other things that the conformal Weyl tensor
$C^{i}\,_{jkl}$ determined from such data vanishes on $\Sigma$. 

There exists a larger class of data which satisfy (i) and (ii) but
not (iii). This raises the questions whether the requirement of
asymptotic simplicity is inappropriately strong. It turns out,
however, that for data violating (iii) the rescaled Weyl tensor  
$d^{i}\,_{jkl}$ diverges in general strongly at $\Sigma$.
Because of the hyperbolicity of the conformal field equations we
can expect that this blow up spreads and leads to a space-time
with a curvature singularity in which the outgoing null geodesics are
not even complete. Questions about their asymptotic behaviour do not
arise for such solutions. Moreover, free data yielding data $h_{\mu
\nu}$, $\chi_{\mu \nu}$ for which (i) and (ii) hold, need only satisfy a
few simple ``extra conditions'' on $\Sigma$ in order for (iii) to be
satisfied. I think it is fair to conclude from what we have seen that:

{\em Asymptotic simplicity is a natural concept for solutions to}
\[
Ric[\tilde{g}] = \lambda\,\tilde{g},\,\,\,\,\,\,\lambda > 0,\,\,\,\,\,
\,\,\,\,\,\,\,sign(\tilde{g}) = (1, -1, -1, -1).
\]

The need for extra conditions on the free data on $\Sigma$ and the fact
that data for the conformal field equations explode if these extra
conditions are not met, may be forboding complications in the case
$\lambda = 0$. In that case ${\cal J}$ should be a null hypersurface.
It may turn out, however, that the singularity ``spreads over the
set which we would like to consider as ${\cal J}$'' and destroys any
hope to find a smooth conformal boundary.

\section{Gravitational fields of isolated systems}

The idea of the conformal boundary was one of the upshots of a discussion
which was initiated by the attempt to arrive at a rigorous concept of
gravitational radiation in the context of gravitational fields of
isolated systems (cf. \cite{friedrich:ted} for a sketch of this
development and a list of relevant references). It is, of course, of
utmost interest to control, qualitatively as well as quantitatively, the
complete process which starts with the generation of the gravitational
radiation by the interactions of some, possibly massive, sources and
which ends with the registration of the gravitational radiation at large
distances from any sources. At present we are far from a position to do
that. However, when I shall concentrate in the following on solutions to
the vacuum equations
\begin{equation}
\label{einst} 
\tilde{R}_{\mu \nu} = 0, 
\end{equation}
it should be kept in mind that the resulting discussion of the asymptotic
behaviour of the gravitational field applies also to situations which
includes black holes or massive and other sources. These are being
ignored here in order not to burden our discussion with a baggage which
is irrelevant for the questions we want to analyse.

Our expectation is, of course, that the asymptotic behaviour of the
gravtitational field of an isolated system resembles that of
Minkowski space. The latter is exhibited most clearly by the well known
conformal extension of Minkowski space (cf. Figure $3$).

\begin{figure}
\centerline{\epsffile{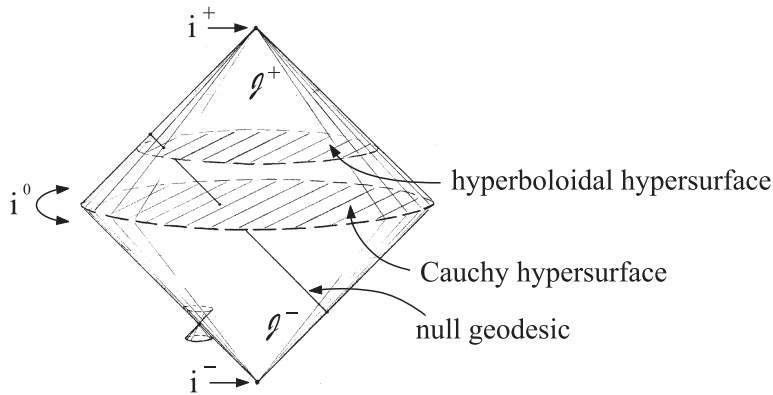}} 
\centering
\begin{minipage}[b]{11cm}
\let \normalsize = \footnotesize
\caption{\sl The conformal extension of Minkowski space.}
\end{minipage}
\end{figure}

Minkowski space satisfies conditions $1$ to $6$ of our definition of
asymptotic simplicity. Moreover, there exist smooth conformal extensions
which contain regular points $i^{\pm}$, representing past and future
time-like infinity, and a regular point $i^0$, which represents
space-like infinity. We shall refer to a solution of the Einstein
equation \ref{einst} arising from asymptotically flat initial data on a
Cauchy hypersurface diffeomorphic to $\Re^3$ and satisfying conditions
$1$ to $6$ as to an ``Minkowski-type space-time''. The occurence of
regular points in the conformal extension corresponding to $i^{\pm}$
will not be required and the occurence of a regular point corresponding
to $i^0$ cannot be required if we are interested in solutions which are
not conformally flat. Nevertheless, we shall follow the custom of
thinking of space-like infinity as being represented by an ideal point
$i^0$. In view of the preceding discussion, one might ask:

{\em Is asymptotic simplicity nonlinearly stable at Minkowski space ?}

It turned out to be much more difficult than expected to find an answer
to this question. In fact, even the ``easier'' question whether there
exist non-trivial Minkowski-type space-times has not been answered
yet. 

A first step into that direction was made by the analysis of the
hyperboloidal initial value problem. Imagine a smooth hypersurface $S$
in the conformal extension of a Minkowski-type space-time which extends
to future null infinity ${\cal J}^{+}$ such that it is space-like
even on the boundary $\Sigma = S \cap {\cal J}^{+}$ of $S$. We call such
a hypersurface together with the initial data induced on it a
``hyperboloidal initial data set'' and the initial value problem for
Einstein's equation \ref{einst} based on these data the ``hyperboloidal
initial value problem''. The following has been shown
\cite{friedrich:Cauchy}, \cite{friedrich:first}.

{\em Solutions to the hyperboloidal initial value problem for
Einstein's equation \ref{einst} are asymptotically simple in the future.
Hyperboloidal initial data sufficiently close to Min\-kows\-kian
hyperboloidal data develop into solutions which admit smooth conformal
extensions containing a regular point $i^{+}$ that represents future
time-like infinity.}

The main point of this result is to reduce the search for
Minkowski-type space-times to the investigation of the solutions in a
neighbourhood of space-like infinity (thinking in terms of the
conformal picture).   

An interesting subtlety arises here. It has been shown in \cite{acf},
\cite{ac} how to construct smooth hyperboloidal initial data from
suitably given ``free data''. Similar to what we have seen in the case
$\lambda > 0$, there exists also in the present case a class of more
general free data which lead to smoothly compactifiable physical
initial data. Again, the free data have to satisfy a few extra
conditions on the boundary $\Sigma$ of $S$ to yield smooth initial data
for the conformal field equations; otherwise the rescaled comformal Weyl
tensor determined by them diverges at $\Sigma$. If such divergent data
develop at all into solutions which are future null complete, they will
certainly not admit a smooth conformal boundary.

Does this indicate the need to impose extra conditions near space-like
infinity if we want to construct Minkowski-type space-times from
asymptotically flat standard Cauchy data ? 

This would be consistent with the results of Christodoulou and
Klainerman \cite{christ:klain}. In their proof of the nonlinear
stability of Minkowski space they could only establish a peeling
behaviour of the conformal Weyl tensor near null infinity which is
weaker than the peeling behaviour implied by asymptotic simplicity. If
their estimates were sharp, this would suggest that asymptotic simplicity
can be established only, if at all, for data satisfying stronger
conditions then those required in \cite{christ:klain}. 

One may wonder about the possible nature of such conditions. 
A straightforward streng\-thening of the usual fall-off conditions of
asymptotically flat Cauchy data would eventually lead to data with
vanishing mass, i.e. to trivial Cauchy data.

\subsection{The basic problem}

Our discussion has indicated the importance to understand the behaviour
of the field in the region where null infinity touches space-like
infinity. To illustrate the situation we take a closer look at
space-like infinity in the standard conformal extension of Minkowski
space (cf. Figure $4$).

\begin{figure}
\centerline{\epsffile{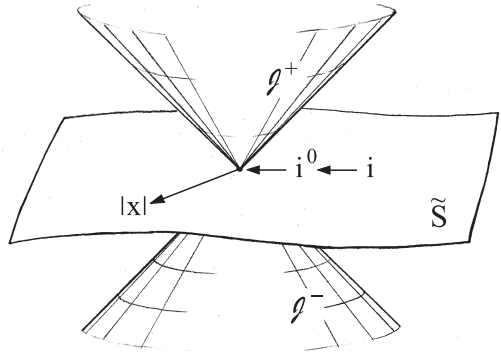}} 
\centering
\begin{minipage}[b]{11cm}
\let \normalsize = \footnotesize
\caption{\sl Neighbourhood of $i^0$ in Minkowski space.}
\end{minipage}
\end{figure}

In this extension the Cauchy hypersurface $\tilde{S} = \{t = 0\}$
(say) of  Minkowski space is compactified by adding of a point
$i$ to obtain a smooth compact hypersurface 
$S = \tilde{S} \cup \{i\} \simeq S^3$. The future (past) null cone at
this point coincides with 
${\cal J}^{+}$ (${\cal J}^{-}$). It will be convenient conceptually
to distinguish the point $i^0$, considered as the endpoint of all
space-like geodesics in the 4-dimensional space, from the point $i$
representing space-like infinity for the asymptotically flat initial
data on $\tilde{S}$. We note that the conformal factor $\Omega$ 
satisfies $\Omega = 0$, $d\,\Omega
\neq 0$ on ${\cal J}^{\pm}$ and $\Omega = 0$, $d\,\Omega = 0$,
$Hess\,\Omega \simeq g$ at $i^0$.   

We could choose now asymptotically flat Cauchy data for Einstein's
equation \ref{einst} on $\tilde{S}$ which are close to the initial data
induced by Minkowski space, transform them appropriately to obtain
the corresponding data for the conformal field equations, and try to
establish the existence of non-trivial Minkowski-type space-times by
using the conformal field equations. Then we encounter the following
difficulty, which poses in fact the main problem of the whole field.
Let $|x(p)|$ denote the distance of the point $p \in S$ from the point
$i$ in terms of the given metric $h_{\mu \nu}$ on $S$. Then one finds
that precisely at the point where the mass of the data manifests itself
the rescaled conformal Weyl tensor behaves like
\[
d^i\,_{jkl} = O(\frac{1}{|x|^3}) 
\quad\mbox{as}\quad |x| \rightarrow 0
\quad\mbox{unless}\quad m_{ADM} = 0. 
\]
This strong divergence at $i$ of some of the initial data for the
conformal field equations has the consequence that the existence of
Minkowski-type solutions cannot be shown by applying straightforward PDE
methods to these equations. 

If one tries to analyse the evolution of the field near $i$ one faces a
host of complicated subproblems. Already the calculation of a formal
expansion of a solution will quickly be swamped by singular terms which
get more and more complicated at any order. Moreover, there are various
questions which have to be analysed in the context of abstract existence
arguments. Just to name a few: (i) How do we choose the gauge ? If the
chosen gauge is propagated implicitly by wave equations the singularity
at $i$ may lead to singular gauge dependent fields, including
$\Omega$, which possibly hide the intrinsic smoothness of the conformal
boundary. (ii) How do we control the conformal life time of the solution
? We would need to ensure that the solutions extend smoothly to
the set ${\cal J} = \{\Omega = 0\}$ but it may already turn out
problematic to locate this set if the gauge dependent function $\Omega$
is non-smooth. Finally, how do we decide whether ``${\cal J}$ is
inherently smooth/non-smooth'' ?  

Our problem is in fact even more complicated. As mentioned above, it
may be necessary to consider a suitably restricted subclass of the Cauchy
data considered e.g. in \cite{christ:klain} to obtain solutions
which admit a smooth structure at null infinity. This poses the
difficult question: How do we relate the smoothness of the solution
near null infinity to properties of the Cauchy data on $S$ and how do
we determine those data which evolve into asymptotically simple
space-times ?

In view of these difficulties a different kind of question arises.
In zeroth order of the conformal structure, i.e. on the level of the
light cone structure or the causal structure, space-like infinity is
represented naturally as an ideal point. For higher order structures
derived from the conformal structure to be smooth this representation is
too narrow. Thus we are led to ask: Does there exist a useful
representation of space-like infinity which is ``finite but wider than
the point
$i^0$'' ?

\subsection{The finite representation of space-like infinity}

It turns out that the investigation of the questions raised above
requires a much deeper analysis of the Cauchy data near
space-like infinity as has been available hitherto. What exactly needs to
be known can only be decided by a simulteneous study of the evolution
problem. In order not to get bogged down by the wealth of details and to
be able to concentrate on the evolution problem I worked out the
following results for a class of data which is rich enough to exhibit
the decisive features of the problem and which, on the other hand,
reduces the amount of algebraic calculations considerably. Furthermore,
I assume the data to be given on 
$\tilde{S} = S \setminus \{i\}$ where $S \simeq S^3$. 
It should be emphasized that this condition and the conditions listed
below are mainly made for convenience. There is ample space for
generalizations and what will be said later on about the
evolution equations is independent of requirements on the data. Our
assumptions are as follows:

{\em The data are time-symmetric, i.e. $\tilde{\chi}_{\mu \nu} = 0$
and the data are specified completely by an asymtotically flat
(negative) Riemannian metric $\tilde{h}_{\mu \nu}$ on $\tilde{S}$ with
vanishing Ricci scalar. The data are ``asymptotically simple'',
i.e. there exists a function $\Omega \in C^{\infty}(\tilde{S}) \cap
C^2(S)$ with 
$\Omega > 0$ on $\tilde{S}$, $\Omega = 0$, $d\,\Omega = 0$, and 
$Hess\,\Omega$ negative definite at $i$ such that 
$h_{\mu \nu} = \Omega^2\,\tilde{h}_{\mu \nu}$ extends to a smooth
Riemannian metric on $S$. In some $h$-normal coordinate system
$x^{\alpha}$ centered in $i$ the metric $h_{\mu \nu}$ is real analytic
close to $i$.}

As a consequence the basic data have the following local expressions near
$i$. The conformal factor has the form
$\Omega = \frac{|x|^2}{(U + |x|\,W)^2}$
where $U$, with $U(i) = 1$, is an analytic function determined by the
local geometry near $i$ and $W$ is an analytic function which encodes
global information since $W(i) = \frac{1}{2}\,m_{ADM}$.
Writing $\Gamma = |x|^2$, the rescaled conformal Weyl tensor takes in
space spinor notation the form 
$\phi_{abcd} = \phi'_{abcd} + \phi^W_{abcd}$ 
with a ``massles part''
\[
\phi'_{abcd} = \Gamma^{-2}\,\left\{ U^2\,D_{(ab}\,D_{cd)}\,\Gamma -
4\,U\,D_{(ab}\Gamma\,D_{cd)}U \right.
\]
\[
\left. - 2\,\Gamma\,U\,D_{(ab}\,D_{cd)}\,U + 6\,\Gamma\,D_{(ab}U\,D_{cd)}U
+ \Gamma\,U^2\,s_{abcd} \right\},
\]
defined in terms of the local geometry near $i$, and a ``massive part''
\[
\phi^W_{abcd} =
\Gamma^{-2}\,
\left\{ - \frac{3}{2}\,\frac{1}{|x|}\,U\,W\,
D_{(ab}\,\Gamma\,D_{cd)}\,\Gamma 
+ U\,W\,|x|\,D_{(ab}\,D_{cd)}\,\Gamma \right. 
\] 
\[
+ 2\,|x|\,(W\,D_{(ab}\Gamma\,D_{cd)}U - 3\,U\,D_{(ab}\Gamma\,D_{cd)}W)
\]
\[
+ 2\,|x|\,\Gamma\,(- U\,D_{(ab}\,D_{cd)}\,W - W\,D_{(ab}\,D_{cd)}\,U
+ 6\,D_{(ab}U\,D_{cd)}\,W + U\,W\,s_{abcd})
\]
\[
\left. + \Gamma^2\,(- 2\,W\,D_{(ab}\,D_{cd)}\,W 
+ 6\,D_{(ab}W\,D_{cd)}\,W
+ W^2\,s_{abcd}) \right\},
\]
where $s_{abcd}$ denotes the trace free part of the Ricci tensor of $h$.
Unless the ADM mass vanishes we still have  
$\phi^W_{abcd} = O(\frac{1}{|x|^3})$ as $|x| \rightarrow 0$.

Analysing under these assumptions the situation near space-like infinity
and letting oneself be guided by the field equations and conformal
geometry, one is led to a representation of space-like infinity which
differs from anything suggested before. The new picture is
indicated in Figure $5$.

\begin{figure}
\centerline{\epsffile{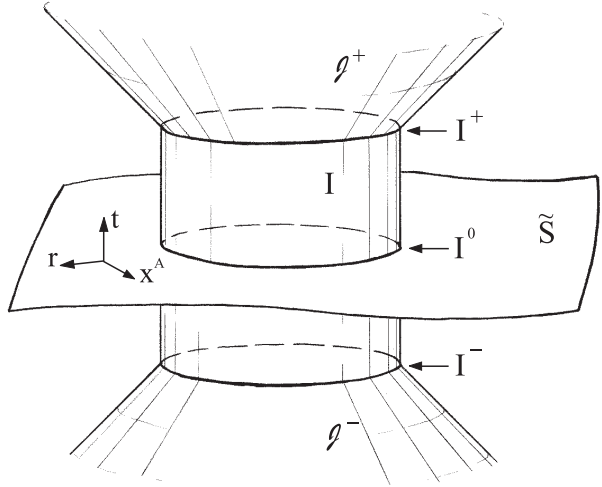}} 
\centering
\begin{minipage}[b]{11cm}
\let \normalsize = \footnotesize
\caption{\sl The cylinder $I$ representing space-like infinity.} 
\end{minipage}
\end{figure}

It is an important element of this setting that the conformal factor,
denoted by $\Theta$ to distiguish it from the function $\Omega$ given on
the initial hypersurface $S$, is known explicitly as a function of the
coordinates and the initial data on $\tilde{S}$. As we shall see, it is
determined completely by the conformal geometry, Einstein's equation,
and certain initial conditions on $\tilde{S}$ which are chosen here
such as to obtain a simple picture. 

The point $i \in S$ is blown up to a spherical set $I^0 \simeq S^2$ which
forms the new boundary of $\tilde{S}$ at space-like infinity. We write
now $S = \tilde{S} \cup I^0$ for the smooth manifold with boundary
$I^0$ so obtained and introduce a coordinate $r$ on $S$ which vanishes on
$I^0$ and is positive elsewhere. 

The coordinate $t$ is chosen such that it vanishes on $S$ and induces an
intrinsic parameter on a congruence of ``conformal geodesics'' orthogonal
to $S$. Then the 4-dimensional space-time $\tilde{M}$ is given near
space-like infinity  in the form $\tilde{M} 
= \{(t, p) \in \Re \times S| r(p) > 0,\,\,\,|t| < 1 + r(p) \}$.
We have $\Theta > 0$ on $\tilde{M}$.

By a natural extension process we obtain the set 
$I = \{(t, p) \in \Re \times S| r(p) = 0,\,\,\,|t| < 1 \}$
which represents space-like infinity. In terms of the earlier picture 
a point of $I$ can be interpreted as a space-like direction at $i^0$,
but $I$ is not defined this way. Furthermore, future and past null
infinity are given near space-like infinity by the sets ${\cal J}^{\pm}
= \{(t, p) \in \Re \times S| r(p) > 0,\,\,\,t = \pm (1 + r(p)) \}$. The
function $\Theta$ vanishes on $I \cup {\cal J}^{-} \cup {\cal J}^{+}$.
The set $I^{\pm} = \{(t, p) \in \Re \times S| r(p) = 0,\,\,\,t = \pm 1
\}$, where ``null infinity touches space-like infinity'', will be of
special interest. In the following I shall indicate the structural
background of the new picture.

\subsection{Conformal geometry and Einstein's equation}

One of the most important features of the new representation is that it is defined
exclusively in terms of the field equations and the conformal geometry or, equivalently,
the light cone structure which in turn determines the physical characteristics of the
field equations. To make full use of the conformal geometry we need to consider conformal
rescalings $\tilde{g} \rightarrow g = \Theta\,\tilde{g}$ 
of the metric as well as transitions from the $\tilde{g}$-Levi-Civita connection to 
``Weyl connections'' 
$\tilde{\nabla} \rightarrow \hat{\nabla} = \tilde{\nabla} + S(b)$, where the
difference tensor, determined by a 1-form $b$, is defined by   
$ S(b)_{\mu}\,^{\nu}\,_{\rho} = 
\delta^{\nu}\,_{\mu}\,b_{\rho} +
\delta^{\nu}\,_{\rho}\,b_{\mu} -
\tilde{g}_{\mu \rho}\,\tilde{g}^{\nu \lambda}\,b_{\lambda}$.
Parallel transport by such connections maps $\tilde{g}$-conformal frames again into
such frames.   

A ``conformal geodesic" \cite{yano} is a space-time curve $x(t)$, together with a
1-form $b(t)$ along it, which satisfy the system of ODE's 
\[ 
(\tilde{\nabla}_{\dot{x}}\dot{x})^{\mu} + 
S(b)_{\nu}\,^{\mu}\,_{\rho}\,\dot{x}^{\nu}\,\dot{x}^{\rho} = 0,\,\,\,\,\,  
(\tilde{\nabla}_{\dot{x}}b)_{\nu} - \frac{1}{2} 
b_{\mu}\,S(b)_{\nu}\,^{\mu}\,_{\rho}\,\dot{x}^{\rho} 
- \tilde{L}_{\nu \mu} \, \dot{x}^{\mu}= 0, 
\]
where
$\tilde{L}_{\nu \mu} = \frac{1}{2}\,(\tilde{R}_{\nu \mu} 
- \frac{1}{6}\,\tilde{g}_{\nu \mu}\,\tilde{R})$ is given in terms of the Ricci
tensor and the Ricci scalar of $\tilde{g}$. 

The 1-form $b(t)$ defines by the formula above a Weyl connection $\hat{\nabla}$ along the
curve $x(t)$ such that the curve is an autoparallel for this connection.  
It is important to note that the curve $x(t)$ and the connection $\hat{\nabla}$ are
invariants of the conformal structure of $\tilde{g}$. 

We construct in an obvious way ``conformal Gauss coordinates based on $S$''
\cite{friedrich:schmidt} which are generated by a congruence of conformal geodesics
threading orthogonally through $S$. The natural parameter $t$ on these curves which
vanishes on $S$ is used as a time coordinate. Beside the connection $\hat{\nabla}$ 
we introduce also a smooth conformal frame field $e_k$ which satisfies
$\hat{\nabla}_{\dot{x}}\,e_k = 0$. This defines in turn a metric $g$ in the conformal
class of $\tilde{g}$ by the requirement 
$g_{ik} \equiv g(e_i, e_k) = diag\,(1, -1, -1, -1)$ and consequently a conformal factor
$\Theta$ satisfying $g = \Theta^2\,\tilde{g}$. 

By this procedure we obtain a ``conformal Gauss gauge'' which is defined
solely in terms of conformal geometry. We study this gauge because we
expect that suitably chosen conformal geodesics starting in the physical space-time will
pass through null infinity into the (prospective smooth) conformal extension without
being affected by the singularity at the point $i$. The freedom to prescribe on $S$
initial data for $\dot{x}$, $b$, $e_k$, whence for $\Theta$, can be used to 
adapt the gauge to the situation we want to study. 

The attempt to combine this gauge with the conformal field equations led to the
following unexpected observation \cite{friedrich:I}. 

{\em Suppose $x(t)$, $b(t)$, $e_k(t)$, $\Theta(t)$ are as above with $t$ taking values
in an interval $J$, $x(t)$ being time-like, and $\dot{x}(t_{*}) = e_0$, 
$\Theta(t_{*}) > 0$ for some $t_{*} \in J$. Then, if Einstein's equation 
$\tilde{R}_{\mu \nu} = 0$ holds on $\tilde{M}$, one has 
\[
\Theta(t) = \Theta_{*} + (t - t_{*})\,\dot{\Theta}_{*} 
+ \frac{1}{2}\,(t - t_{*})^2\,\ddot{\Theta}_{*},\,\,\,\,\,\,\,
d_k(t) \equiv \Theta\,b_{\mu}\,e^{\mu}\,_k = 
(\dot{\Theta}, d_{a *})
\]
with constants $\Theta_{*}$, $\dot{\Theta}_{*}$, $\ddot{\Theta}_{*}$,
$d_{a *}$, $a = 1, 2, 3$.} 

Thus the conformal factor $\Theta$ is known explicitly as a function of the
parameter $t$ and certain data at $t_{*}$. Moreover, it is quadratic in $t$ 
and, provided the underlying conformal geodesic has a sufficiently long life-time,
we can expect that there are numbers $t_{\pm} \in J$, $t_{-} < t_{*} < t_{+}$ with 
$\Theta(t_{\pm}) = 0$. This is precisely what would happen if $x(t)$ would pass the
conformal boundary at $t_{\pm}$, i.e. if we had $x(t_{\pm}) \in {\cal J}^{\pm}$.

There is some information about Einstein's equation \ref{einst} encoded in the 
expressions for $\Theta(t)$, $d_k(t)$. Under the assumption that the congruence of
conformal geodesics passes smoothly through ${\cal J}$, the expressions imply
$g^{ij}\,\nabla_{i}(\Theta)\,\nabla_{j}(\Theta) = 0$ on ${\cal J}$, a relation
which was deduced before from the field equations.

To incorporate the conformal Gauss gauge into the field equations we need to
extend the conformal field equations to admit Weyl connections. The ``generalized
conformal Einstein equations'' so obtained form a system of equations for the unknown 
\[
u = \left(e^{\mu}\,_k,\,
\hat{\Gamma}_i\,^j\,_k,\,
\hat{L}_{jk} = 
\frac{1}{2}\,\hat{R}_{(jk)} - \frac{1}{12}\,\hat{R}\,g_{jk}
- \frac{1}{4}\,\hat{R}_{[jk]},\,
d^i\,_{jkl} = \Theta^{-1}\,C^i_{jkl}\right),
\]
where $\hat{\Gamma}_i\,^j\,_k$ are the connections coefficients of the Weyl connection 
$\hat{\nabla}$ in the frame $e_k$, $\hat{R}_{jk}$ is the Ricci tensor of
$\hat{\nabla}$, and $\hat{R}$ = $g^{jk}\,\hat{R}_{jk}$. The system is given by  
\[
[e_{p},e_{q}] =
(\hat{\Gamma}_{p}\,^{l}\,_{q} - \hat{\Gamma}_{q}\,^{l}\,_{p})\,e_{l},
\]
\[
e_{p}(\hat{\Gamma}_{q}\,^{i}\,_{j}) - 
e_{q}(\hat{\Gamma}_{p}\,^{i}\,_{j}) 
- 2\,\hat{\Gamma}_{k}\,^{i}\,_{j}\,\hat{\Gamma}_{[p}\,^{k}\,_{q]}
+ 2\,\hat{\Gamma}_{[p}\,^{i}\,_{|k|}\,\hat{\Gamma}_{q]}\,^{k}\,_{j} 
\]
\[
= 2\,g^{i}\,_{[p}\,\hat{L}_{q]j}
- 2\,g^{ik}\,g_{j[p}\,\hat{L}_{q]k}
- 2\,g^{i}\,_{j}\,\hat{L}_{[pq]}
+ \Theta\,d^{i}\,_{jpq},  
\]
\[
\nabla_{i}\,d^{i}\,_{jkl} = 0,
\]
\[
\hat{\nabla}_{p}\,\hat{L}_{qj} 
- \hat{\nabla}_{q}\,\hat{L}_{pj} 
= d_{i}\,d^{i}\,_{jpq}.
\]
We note that no differential equations for the fields $\Theta$, $d_k$ are included. For
our purpose the most important property of this system is stated in the following result
\cite{friedrich:I}.  

{ \em In the conformal Gauss gauge, with $\Theta(t)$, $d_k(t)$ being the explicit
expressions above, the generalized conformal Einstein equations imply a ``reduced
system'' of evolution equations which (i) is symmetric hyperbolic irrespective of the
values taken by $\Theta$, (ii) preserves the constraints under its evolution, (iii) gives
for initial data satisfying the constraints solutions $\tilde{g}_{\mu \nu} =
\Theta^{-2}\,g_{\mu \nu}$ to Einstein's equation \ref{einst} in the region where $\Theta
> 0$.}

The conformal factor $\Theta$ and the 1-form $d_k$ play the role of gauge source
functions in the equations (cf. \cite{friedrich:hypred}). The propagation equations for
the fields $e^{\mu}\,_k$, $\hat{\Gamma}_i\,^j\,_k$, $\hat{L}_{jk}$ contain only the
derivative operator $\partial_t$. This simplifies the analysis near $i$
considerably and clearly exhibits the special status of the Bianchi equation.

Given the equations above, it turns out that one can adapt the conformal Gauss gauge in a
particularly convenient way to the situation near space-like infinity. All the details
which are to be observed here and also the following results can be found in
\cite{friedrich:II}. I mention just one basic choice. We set 
$\Theta = \Theta_{*} \equiv \kappa^{-1}\,\Omega$, 
$b = \Omega^{-1}\,d\,\Omega$ on $\tilde{S}$ with
$\kappa = r\,\kappa'$ where $\kappa'$ is smooth and satisfies
$\kappa'(i) = 1$. Here $r(p)$ denotes the $h$-distance of a point 
$p \in S$ from $i$. Assuming in addition $d_{\mu}\,e^{\mu}\,_0 = 0$ on
$\tilde{S}$, the fields $\Theta$, $b$ are determined on $\tilde{M}$ and
its possibly smooth conformal extensions. We note that the conformal
gauge on $\tilde{M}$ given by $\Theta$ is related in a particular way to
the conformal gauge of the initial data specified by $\Omega$. 

We perform now the blow up by which the point $i$ is replaced by the sphere $I^0$ and use
$r$ to define a coordinate, again denoted by $r$, which vanishes on $I^0$ and is
positive on $S \setminus I^0$. We shall consider below an extension of $S$ beyond $I^0$
which is defined by allowing $r$ to take negative values. When the complete set of
initial data for the generalized conformal Einstein equations is transferred
appropriately to the new setting we find the following surprising result.  

{\em The function $\kappa'$ can be chosen such that 
$\Theta = \Theta_{*}\,\left(1 - \frac{t^2}{(1 + r)^2}\right)$, with $\Theta_{*}$
constant along the conformal geodesics, and such that $\Theta$, $d_k$ can be 
extended smoothly into a region where $r \le 0$. Moreover, the initial data for the
generalized conformal Einstein equations can be extended smoothly into a domain
where $r \le 0$.}

This observation allows us to set up a {\em regular, finite initial value problem near
space-like infinity} which gives the new picture of space-like infinity outlined
above. The set $I$ representing space-like infinity is obtained in our gauge by a simple
smooth extension and, if the solution extends smoothly to the set $\{t = \pm (1 + r)\}$,
a conformal boundary ${\cal J}$ is obtained by a smooth extension as well. 

\subsection{The cylinder $I$ at space-like infinity}

In the following I want to discuss the nature of the finite, regular 
initial value problem at space-like infinity and some of the consequences
which have been worked out so far \cite{friedrich:II}. Since it is
defined exclusively in terms of the conformal geometry, this initial value
problem ``is always there''. We can ignore it but we cannot avoid its
consequences. Any results derived from it gives direct information on the
conformal geometry.

Since the data and the equations extend smoothly into the set $\{r \le 0\}$, 
we get on the closure of $\tilde{M}$ a problem for symmetric hyperbolic equations
of the form 
\[
(A^t\,\partial_t + A^r\,\partial_r + A^B\,\partial_B)\,u = F(t, r, x^A; u),
\] 
with matrix valued functions $A^{\mu}$ of the unknown ``vector''$u$ and the
coordinates $x^0 = t$, $x^1 = r$, $x^A$, $A = 1, 2$, where the latter imply
in particular local coordinates on the sphere $I^0$. 

A curious property of the new problem is the occurrence of the ``boundary''
$I$. To understand its nature we extend the initial data and the equations
smoothly into the region $\{r \le 0\}$ and use the well known results on
symmetric hyperbolic equations (cf. \cite{kato}, \cite{taylor:123}) to
establish the existence of a smooth solution in some neighbourhood $U$ of
the extended initial hypersurface $S$. Obviously $U$ includes a part of $I$.
Given this solution, we can conclude from the form of the equations and the
data that

{\em The cylinder $I$ at space-like infinity is ``totally characteristic''
in the sense that $A^r = 0$ on $I$}.

Restricting our propagation equations to $I$ we thus get interior
symmetric hyperbolic equations
$(A^t\,\partial_t + A^B\,\partial_B)\,u = F(t, 0, x^A; u)$ 
on $I$ which allow us to determine $u$ on $I$ uniquely in terms of the
values of $u$ on $I^0$. It follows, as expected, that the solution is
determined on $\tilde{M} \cap U$ uniquely by the data on $\tilde{S}$. The
extension of our original problem into the range where $r < 0$ is just a
convenient trick to establish the existence of a solution which
extends smoothly to $I \cap U$.

For $p = 0, 1, 2, \ldots$ we denote by $u_p = \partial^p_r\,u|_I$ the
restriction of the $p$-th $r$-derivative of $u$ to $I$. Taking formal
$r$-derivatives of our original propagation equations and restricting to
$I$, we find

{\em The quantities $u_p$, $p = 1, 2, \ldots$, can be determined
on $I$ recursively as solutions to linear symmetric hyperbolic equations on
$I$. The formal expansion 
$u = \sum_{p \ge 0}\frac{1}{p!}\,u_p(t, x^A)\,r^p$ of $u$ defined by
these coefficients on $I$ is convergent in a neighbourhood of $I^0$.}

We can calculate recursively explicit formulae for the quantities $u_p$.
Though the expressions become more and more complicated for increasing
$p$ and we are still far from having worked out all the details we can
state the most important consequence of the observation above.

{\em The interior equations on $I$ allow us to relate properties of the solution
on ${\cal J}^{\pm}$ near $I^{\pm}$ to the structure of the data on $S$ near
$I^0$ if the solution extends smoothly to ${\cal J}^{\pm}$.}  

As an example, J. K\'ann\'ar \cite{kannar:private} has derived (observing
the regularity condition given below) formulae for the conserved
quantities of Newman and Penrose \cite{newman:penrose} in terms of the
functions $u_p$, $p = 0,\ldots,3$. Another example is given by the following
result. Since the total characteristic $I$ approaches the characteristic
${\cal J}^{\pm}$ transversely in $I^{\pm}$ we can expect a degeneracy of
the equations on this set. The calculation of $u$ on $I$ shows in fact
that the matrix $A^t$, which is positive definite on $I$, degenerates on
$I^{\pm}$. As a  consequence the functions $u_p$ develop a certain type of
logarithmic singularity on $I^{\pm}$. For a correct assessment of this
observation it is important that our setting is defined completely in terms
of the conformal geometry and not based on some obscure gauge condition.
The observed singularities indicate an intrinsic property of the conformal
structure. It turns out that we can isolate those parts of the initial data
which give rise to the singular terms. 

{\em The logarithmic singularities alluded to above do not occur if $h$ 
satisfies at the point $i$ the regularity condition
\[
{\cal C}(D_{i_1}, \dots, D_{i_p}\,B_{jk})(i) = 0,\,\,\,\,\,
p = 0, 1, 2, \ldots, 
\]
where $B_{jk}$ denotes the Cotton tensor of $h$ and ${\cal C}$ means 
``symmetrize and take the trace free part''. This regularity condition is
equivalent to the requirement that the massless part $\phi'_{abcd}$ of
the rescaled conformal Weyl tensor extend smoothly to the point $i$ of
the initial hypersurface.} 

The regularity condition is conformally invariant and thus in fact a condition on
the free initial data. Complicated as it looks, the condition is in fact quite weak
and appears to impose hardly any restriction on the modeling of physically
interesting systems. It allows us to choose the free data arbitrarily on
any given compact subset of $\tilde{S}$, provided we observe, as usual,
the solvability of the Lichnerowicz equation. Furthermore, the condition
is satisfied by all static solutions which are analytic at space-like
infinity \cite{beig:confstat},
\cite{friedrich:static}. Thus it implies no restriction on the
admissible multipoles.

We note, that the set $I$ is obtained as a limit of conformal geodesics
and that its points can be interpreted as the ``space-like directions at 
$i^0$''. Apart from that it does not have any geometric meaning. The
light cone  structure degenerates on $I$ but the vector fields used to
express the equations extend smoothly to $I$. We find that
$g(\partial_r, \partial_r)(p) \rightarrow \infty$ as $p \rightarrow I$,
$|t(p)| < 1 - \epsilon$, $0 < \epsilon <1$. If we replace $r$ by a 
radial coordinate $\rho$ which is better adapted to the conformal gauge
defined by $\Theta$ in the sense that 
$0 < c < g(\partial_{\rho}, \partial_{\rho}) < c' < \infty$ we find that
$\rho \rightarrow \infty$ as $r \rightarrow 0$. Thus

{\em In terms of coordinates $\rho$, $t$ adapted to the conformal gauge defined by
$\Theta$, the set $I$ is at $\rho = \infty$ but ``has finite circumference''. The
set ${\cal J}^{\pm}$  is finite in time $t$.}

It is a remarkable fact that the field $d_k$ and the coordinate $r$,
which are adapted to the conformal factor $\Omega$, conspire with
the conformal gauge defined by $\Theta$ to provide a smooth finite 
representation of space-like infinity. We thus arrived at the 
picture of an isolated system indicated in Figure $6$.

\begin{figure}
\centerline{\epsffile{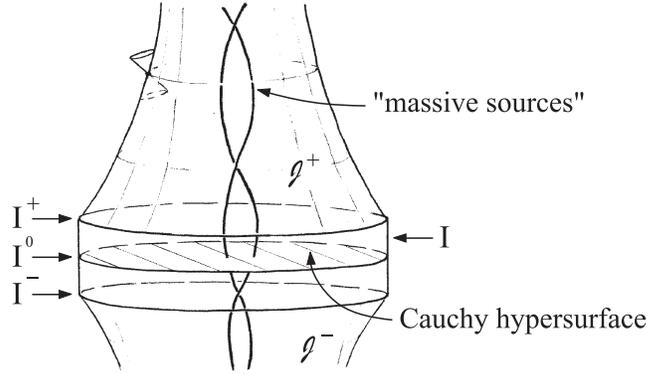}} 
\centering
\begin{minipage}[b]{11cm}
\let \normalsize = \footnotesize
\caption{\sl An isolated gravitating system in the new representation.}
\end{minipage}
\end{figure}

A detailed comparison of the analysis of space-like infinity outlined 
in this lecture with other investigations of space-like infinity  (cf.
\cite{ashtekar:1984}, \cite{ashtekar:romano}, \cite{beig:schmidt},
\cite{christ:klain}, \cite{chrusciel:spi II}, \cite{sommers:1978}) is
impossible here. Naturally, all these studies have various aspects in
common. The present work differs from the work quoted above in that it
combines the following features in one approach. The picture of
space-like infinity which is proposed here is based on assumptions on
the initial data, the structure of the field equations, and properties
of conformal geometry. No a priori assumptions on the time evolution are
being made. It is designed as a basis for an existence theorem which
should enable us to derive statements about the smoothness of the
structure at null infinity. This theorem has still to be worked out. If
it will be done, the setting will allow us to analyse the consequences
of the field equations near space-like infinity to an arbitrary
precision and we shall be able to relate qualitative and quantitative
properties of the fields on ${\cal J}$ near space-like infinity to the
structure of the initial data. Moreover, the setting has the important
practical feature that the evolution of the fields can be studied in a
finite picture near space-like and null infinity with the usual concepts
of smoothness.

\section{Concluding Remarks}

The conformal structure proved to be most important in deriving  -- on
the level of differential topology and geometry -- statements on the
behaviour of space-times in the large. The discussion above shows
clearly that it is of structural interest to understand the
interaction between conformal geometry and Einstein propagation in
depth. Our statements above about the existence and the behaviour of
solutions in the large were obtained by results concerning this
relation. There is still much more to be learned about it.  

Though it has not been emphasized above, certain results about
the hyperbolicity of the Einstein equations form an important technical
ingredient in our analysis. Since the conformal field equations contain
the Bianchi identity as a central part, hyperbolic reductions of the
Einstein equations based on equations for the curvature tensor play a
vital role in our discussions. Since the Bianchi identity is a tensor
equation, the hyperbolicity of the propagation equations derived from it
is independent of the gauge conditions imposed on the lower order fields
and the formalism used (frame formalism, spin frame formalism, ADM
representation of the metric etc.). There is a considerable variety of
possibilities to impose gauge conditions such that the coupled system
governing the curvature and the lower oder fields is symmetric
hyperbolic. These reductions can conveniently be adapted to
characteristic initial value problems, Cauchy problems, and initial
boundary value problems (cf. \cite{friedrich:hypred} and the references
given there). Their applicability is not restricted to the vacuum or the
conformal vacuum equations, they can be extended to include matter
fields. Recently they have been used for the coupled Einstein-Euler
equations to combine symmetric hyperbolicity with a Lagrangian
representation of the flow field \cite{friedrich:pfl}.

We have seen how to derive by straightforward calculations on 
the cylinder at space-like infinity information about the asymptotic
behaviour of solutions in the region where null infinity touches
space-like infinity. However, the fact that infinite problems for
Einstein's equations can be converted into regular finite problems for
the conformal field equations is of even more practical interest. 
It offers the possibility to calculate numerically entire asymptotically
flat space-times together with their conformal boundary on finite grids.
The introduction of cut-offs of the field at artificial time-like
boundaries is not necessary. By a judicious choice of the space-like
slicing in the neighbourhood of the cylinder $I$ it should be possible
to perform a smooth transition from the standard Cauchy problem for the
conformal field equations to the hyperboloidal Cauchy problem. 

Numerical calculations based on the hyperboloidal Cauchy problem have
been performed successfully by H\"ubner \cite{hubner:1993},
\cite{hubner:1995}, \cite{hubner:1996}, \cite{hubner:1998} and
Frauendiener \cite{frauendiener:1997a}, \cite{frauendiener:1997b}. 
The full potential of hyperboloidal hypersurfaces, which offer a convenient way to
trace radiation in the context of Cauchy problems, can only 
completely be exhausted in conjuction with the conformal field
equations. The radiation extraction is performed directly at null
infinity where a well defined concept of a ``radiation field'' is
available \cite{penrose:let}, \cite{penrose:scri}.

The fact that the requirement of smoothness of the fields near $I^{\pm}$
implies regularity conditions on the data raises new questions about the
concept of an ``isolated system'' in general relativity (cf. also
\cite{friedrich:rp} for a discussion of this point). Several authors
(cf. \cite{cms}, \cite{winicour:log}) considered asymptotic expansions
of the gravitational field near null infinity which admit logarithmic
terms. I think that one should not give in so easily. We have seen that
the somewhat esoteric desire to get control on the asymptotic smoothness
of solutions yields results of immediate practical consequences.
Moreover, the idea of an ``isolated system'' is an idealization which
introduces ``space-like infinity'' as a convenient construct. These
notions leave a certain freedom which should be exploited as
far as necessary but without introducing irrelevant information. If in
the end it should turn out that the requirement of asymptotic simplicity
restricts the class of admissible solutions too strongly to model
certain situations of physical interest, we will have understood why
there is a need to generalize and can start to do so. 

Perhaps also Ellis's critique \cite{ellis:1984} of the notion of an
isolated system considered in this lecture should be mentioned here. It
was argued in \cite{ellis:1984} that one should avoid this idealization,
seperate instead the system of interest (star, system of stars etc.)
from the rest of the universe by a cut along a time-like hypersurface
and study the system so obtained. This introduces an initial boundary
value problem into the discussion of the system. Apart from the fact
that it is hard to see how we could get useful information on the
required boundary data, this problem introduces difficulties of its own
\cite{friedrich:nagy}. In view of this and the results discussed above
it appears to me that at least the technical aspects of the critique in
\cite{ellis:1984} need to be reconsidered in the light of the recent
developments.


\begin{thebibliography}{11}

\bibitem{acf}
L. Andersson, P.T. Chru\'sciel, H. Friedrich.
\newblock On the regularity of solutions to the Yamabe equation and the
existence of smooth hyperboloidal initial data for Einstein's field
equations. 
\newblock {\em Commun. Math. Phys.}, 149 (1992) 587 - 612.

\bibitem{ac}
L. Andersson, P.T. Chru\'sciel.
\newblock On ``Hyperboloidal'' Cauchy Data for Vacuum Einstein Equations
and Obstruction to the Smoothness of Scri.
\newblock {\em Commun. Math. Phys.}, 161 (1994) 533 - 568.

\bibitem{ashtekar:1984}
A. Ashtekar.
\newblock Asymptotic properties of isolated systems: Recent developments.
\newblock In: {\em General Relativity and Gravitation}
\newblock B. Bertotti et. al (eds.), Reidel, Dordrecht, 1984.

\bibitem{ashtekar:romano}
A. Ashtekar, J.D. Romano.
\newblock Spatial infinity as a boundary of space-time.
\newblock {\em Class. Quantum Grav.}, 9 (1992) 1069 - 1100.

\bibitem{beig:confstat}
R. Beig.
\newblock Conformal properties of static space-times. 
\newblock {\em Class. Quantum Grav.} 8 (1991) 263 - 271.

\bibitem{beig:schmidt}
R. Beig, B.G. Schmidt.
\newblock Einstein's Equations near Spatial Infinity.
\newblock {\em Commun. Math. Phys.}, 87 (1982) 65 - 80. 

\bibitem{bicak}
J. Bi\v{c}\'ak.
\newblock Radiative spacetimes: Exact approaches.
\newblock In: {\em Relativistic Gravitation and Gravitational Radiation}
\newblock J.-A. Marck, J.-P. Lasota (eds.), Cambridge University Press 1997.

\bibitem{christ:klain}
D. Christodoulou, S. Klainerman.
\newblock The Global Nonlinear Stability of the Minkowski Space.
\newblock Princeton University Press, Princeton 1993.

\bibitem{chrusciel:spi II}
P.T. Chru\'sciel.
\newblock On the Structure of Spatial Infinity: II. Geodesically Regular
Ashtekar-Hansen Structures. 
\newblock {\em J. Math. Phys.} 30 (1989) 2094 - 2100. 

\bibitem{cms}
P.T. Chru\'sciel, M. A. H. MacCallum, D. B. Singleton.
\newblock Gravitational Waves in General Relativity. XIV: Bondi Expansions and
the ``Polyhomogeneity'' of $Sri$.
\newblock {\em Phil. Trans. Royal Soc, London}, A 350 (1995) 113 - 141.

\bibitem{ellis:1984}
G. F. R. Ellis.
\newblock Relativistic Cosmology: Its Nature, Aims, and Problems.
\newblock In: {\em General Relativity and Gravitation}
\newblock B. Bertotti et. al (eds.), Reidel, Dordrecht, 1984.

\bibitem{frauendiener:1997a}
J. Frauendiener.
\newblock Numerical treatment of the hyperboloidal initial value problem for
the vacuum Einstein equations I. The conformal field equations.
\newblock {\em gr-qc/9712050} 

\bibitem{frauendiener:1997b}
J. Frauendiener.
\newblock Numerical treatment of the hyperboloidal initial value problem for
the vacuum Einstein equations II. The evolution equations. 
\newblock {\em gr-qc/9712052} 

\bibitem{friedrich:1981a}
H. Friedrich.
\newblock On the regular and the asymptotic characteristic initial value
problem for Einstein's vacuum field equations. 
\newblock In: {\em Proceedings of the Third Gregynoc Relativity Workshop}
\newblock M. Walker (ed.), Max-Planck Green Report MPI-PAE/Astro 204,
(1979),
\newblock {\em Proc. Roy. Soc.}, A 375 (1981) 169 - 184.

\bibitem{friedrich:1981b}
H. Friedrich.
\newblock The asymptotic characteristic initial value problem for Einstein's
vacuum field equations as an initial value problem for a first-order
quasilinear symmetric hyperbolic system. 
\newblock {\em Proc. Roy. Soc.}, A 378 (1981) 401 - 421.

\bibitem{friedrich:Cauchy}
H. Friedrich.
\newblock Cauchy Problems for the Conformal Vacuum Field Einstein in General
Relativity.
\newblock {\em Commun. Math. Phys.}, 91 (1983) 445 - 472.

\bibitem{friedrich:first}
H. Friedrich.
\newblock On the existence of n-geodesically complete or future complete
solutions of Einstein's field equations with smooth asymptotic structure.
\newblock {\em Commun. Math. Phys.}, 107 (1986) 587 - 609.

\bibitem{friedrich:static}
H. Friedrich.
\newblock On static and radiative space-times.
\newblock {\em Commun. Math. Phys.}, 119 (1988) 51 - 73.

\bibitem{friedrich:global}
H. Friedrich.
\newblock On the global existence and the asymptotic behaviour
of solutions to the Einstein-Maxwell-Yang-Mills equations.
\newblock {\em J. Diff. Geom.}, 34 (1991) 275 - 345. 

\bibitem{friedrich:ted}
H. Friedrich.
\newblock Asymptotic Structure of Space-Time.
\newblock In: {\em Recent Advances in General Relativity}
\newblock A. I. Janis, J. R. Porter (eds.), Birkh\"auser, Basel, 1992.

\bibitem{friedrich:I}
H. Friedrich.
\newblock Einstein Equations and Conformal Structure: Existence of
Anti-de Sitter-Type Space-Times.
\newblock {\em J. Geom. Phys.}, 17 (1995) 125 - 184.

\bibitem{friedrich:hypred}
H. Friedrich.
\newblock Hyperbolic reductions for Einstein's equations.
\newblock {\em Class. Quantum Gravity}, 13 (1996) 1451 - 1469.

\bibitem{friedrich:II}
H. Friedrich.
\newblock Gravitational fields near space-like and null infinity.
\newblock {\em J. Geom. Phys.}, 24 (1998) 83 - 163.  

\bibitem{friedrich:rp}
H. Friedrich.
\newblock Einstein Equations and Conformal Structure.
\newblock In: {\em The Geometric Universe: Science, Geometry and the Work of
Roger Penrose}
\newblock S. A. Hugget et al. (eds.), Oxford University Press, Oxford, 1998.

\bibitem{friedrich:pfl}
H. Friedrich.
\newblock Evolution equations for gravitating ideal fluid bodies in
general relativity.
\newblock {\em Phys. Rev. D}, 57 (1998) 2317 - 2322.

\bibitem{friedrich:nagy}
H. Friedrich, G. Nagy.
\newblock The initial boundary value problem for Einstein's vacuum field
equation.
\newblock {\em Preprint} (1998).

\bibitem{friedrich:schmidt}
H. Friedrich, B.G. Schmidt.
\newblock Conformal geodesics in general relativity.
\newblock {\em Proc. R. Soc. Lond.}, A 414 (1987) 171 - 195. 

\bibitem{geroch}
R. Geroch.
\newblock Asymptotic Structure of Space-Time.
\newblock In: {\em Asymptotic Structure of Space-Time}
\newblock B. F. P. Esposito, L. Witten (eds.), Plenum Press, New York, 1977.

\bibitem{hubner:1993}
P. H\"ubner.
\newblock Numerische und analytische Untersuchungen von (singul\"aren)
asymp\-totisch flachen Raumzeiten mit konformen Techniken. 
\newblock {\em Thesis, Universit\"at M\"unchen} 1993.

\bibitem{hubner:1995}
P. H\"ubner.
\newblock General relativistic scalar-field models and asymptotic flatness.
\newblock {\em Class. Quantum Grav.} 12 (1995) 791 - 808.

\bibitem{hubner:1996}
P. H\"ubner.
\newblock Method for calculating the global structure of (singular) spacetimes.
\newblock {\em Phys. Rev. D} 53 (1996) 701.

\bibitem{hubner:1998}
P. H\"ubner.
\newblock Black hole space-times on grids with trivial boundaries.
\newblock {\em Preprint} 

\bibitem{kannar}
J. K\'ann\'ar.
\newblock Hyperboloidal initial data for the vacuum Einstein equations with
cosmological constant.
\newblock {\em Class. Quantum Grav.} 13 (1996) 3075 - 3084.

\bibitem{kannar:private}
J. K\'ann\'ar.
\newblock Private communication (1997).

\bibitem{kato}
T. Kato.
\newblock The Cauchy problem for quasi-linear symmetric hyperbolic systems.
\newblock {\em Arch. Ration. Mech. Anal.}, 58 (1975) 181 - 205.

\bibitem{newman:penrose:I}
E. T. Newman, R. Penrose.
\newblock An Approach to Gravitational Radiation by a Method of Spin
Coefficients.
\newblock {\em J. Math. Phys.} 3 (1992) 566 - 578.

\bibitem{newman:penrose}
E. T. Newman, R. Penrose.
\newblock New Conservation Laws for Zero-Rest-Mass Fields in Asymptotically
Flat Space-Time. 
\newblock {\em Proc. Roy. Soc.} A 305 (1968) 175 - 204.

\bibitem{penrose:let}
R. Penrose.
\newblock Asymptotic properties of fields and space-time.
\newblock {\em Phys. Rev. Lett.}, 10 (1963) 66 - 68.

\bibitem{penrose:scri}
R. Penrose.
\newblock Zero rest-mass fields including gravitation: asymptotic behaviour.
\newblock {\em Proc. Roy. Soc. Lond.}, A 284 (1965) 159 - 203.

\bibitem{sommers:1978}
P. Sommers.
\newblock The Geometry of the Gravitational Field at Space-Like Infinity.
\newblock {\em J. Math. Phys.} 19 (1978) 549 - 554.

\bibitem{taylor:123}
M. E. Taylor.
\newblock Partial Differential Equations, Vol. 3.
\newblock Springer, New York, 1996.

\bibitem{winicour:log}
J. Winicour.
\newblock Logarithmic asymptotic flatness.
\newblock {\em Found. of Phys.} 15 (1985) 605 - 616.

\bibitem{yano}
K. Yano.
\newblock Sur le circonf\'erences g\'en\'eralis\'ees dans les espace \`a
connexion conforme.
\newblock {\em Proc. Imp. Acad. Tokyo}, 14 (1938) 329 - 332.
\newblock Sur la th\'eorie des espaces \`a connexion conforme.
\newblock {\em J. Fac. Sci. Univ. Tokyo}, Sect. 1, 4 (1939) 1 - 59.

\end{thebibliography}
\end{document}